\documentclass[sigconf]{acmart}

\usepackage[english]{babel}
\usepackage{blindtext}

\usepackage{graphicx}
\usepackage{subcaption}

\usepackage{array}

\usepackage{url}

\usepackage{balance}

\usepackage{listings}

\usepackage{simplebnf}

\usepackage[ruled,vlined,linesnumbered]{algorithm2e}
\usepackage[noend]{algorithmic}

\usepackage{tikz}
\usetikzlibrary{automata, arrows.meta, positioning}

\usepackage{cleveref}
\crefformat{section}{\S#2#1#3}
\Crefformat{section}{\S#2#1#3}

\renewcommand\footnotetextcopyrightpermission[1]{} 
\setcopyright{none}

\acmDOI{}

\acmISBN{}

\acmPrice{}

\newcommand{\sysname}{\textrm{Rela}\xspace}
\newcommand{\flowscale}{6}
\newcommand{\Zone}{Zone\xspace}
\newcommand{\zone}{zone\xspace}

\newcommand{\IE}{\emph{i.e.}\xspace}

\newcommand{\paraspace}{\vspace{1ex}}
\newcommand{\para}[1]{\paraspace\noindent\textbf{#1.}\xspace}

\newcommand{\todo}[1]{\textcolor{blue}{[#1]}}

\newcommand{\zak}[1]{\textcolor{brown}{[Zak: #1]}}
\newcommand{\dpw}[1]{\textcolor{blue}{[dpw: #1]}}
\newcommand{\OMIT}[1]{}

\lstdefinelanguage{DSL}{
  keywords={define, locations, drop, preserve, add, remove, replace, any, else, spec, regex, union, select, where, pspec},
  keywordstyle=\color{black}\bfseries,
  comment=[l]{//},
  commentstyle=\color{green}\itshape,
}

\begin{document}
\title{Relational Network Verification}


\author{
\rm{Xieyang Xu$^{\text{1}}$ \enskip 
Yifei Yuan$^{\text{2}}$ \enskip 
Zachary Kincaid$^{\text{3}}$ \enskip 
Arvind Krishnamurthy$^{\text{1}}$ \enskip \\
Ratul Mahajan$^{\text{1}}$ \enskip 
David Walker$^{\text{3}}$ \enskip
Ennan Zhai$^{\text{2}}$ 
}}
\affiliation{
  \institution{$^{\text{1}}$University of Washington\enskip \enskip \enskip $^{\text{2}}$Alibaba Cloud\enskip \enskip \enskip
$^{\text{3}}$Princeton University}
}



\begin{abstract}
Relational network verification is a new approach to validating network changes. In contrast to traditional network verification, which analyzes specifications for a single network snapshot, relational network verification analyzes specifications concerning two network snapshots (e.g., pre- and post-change snapshots) and captures their similarities and differences. 
Relational change specifications are compact and precise because they specify the flows or paths that change between snapshots and then simply mandate that other behaviors of the network ``stay the same'', without enumerating them.  To achieve similar guarantees, single-snapshot specifications need to enumerate all flow and path behaviors that are not expected to change, so we can check that nothing has accidentally changed. Thus, precise single-snapshot specifications are proportional to network size, which makes them impractical to generate for many real-world networks.

To demonstrate the value of relational reasoning, we develop a high-level 
relational specification language and a tool called \sysname to validate network changes. \sysname first compiles input specifications and network snapshot representations to finite state transducers. It then checks compliance using decision procedures for automaton equivalence. Our experiments using data on complex changes to a global backbone (with over $10^3$ routers) find that \sysname specifications need fewer than 10 terms for 93\% of them and it validates 80\% of them within 20 minutes.

\end{abstract}

\maketitle

\section{Introduction}
\label{sec:intro}



One of the riskiest network management activities today is changing a running network. Outages can occur during changes because of incorrect change implementation (e.g., accidentally blocking traffic) or latent bugs (e.g., traffic starts traversing a longstanding filter). When changes go wrong, banks go offline, airlines stop flying, emergency services become unreachable, and businesses lose millions of dollars~\cite{time-warner,united,twitter,bank,t-mobile,cloudflaremisconfiguration,azuremisconfiguration}. Since changing a network to alter its security posture, optimize resource usage, or add capacity is unavoidable, we must make changes safer to make networks more reliable.

The last decade has seen remarkable progress toward verification technologies that can reason about large, real-world networks. These technologies typically tell a user whether a \emph{single} network snapshot $N$
satisfies specification $S$.  The snapshot may be for an updated network configuration that engineers
wish to deploy, and the specification 
may demand that DNS traffic is never blocked or no packet can reach the high-security zone without traversing the firewall.
Indeed, many large networks use these technologies today~\cite{batfish-2023,rcdc,aws-inspector,libra}. 

\emph{Single-snapshot verification} tools, while valuable, do not suffice for keeping networks running reliably as they are updated.  Consider a common network change that moves all traffic on link A to link B as a precursor to shutting A for maintenance. The engineer wants to ensure that all traffic on link A is moved, it is moved to link B and nowhere else, and that no other traffic is impacted.  To use single-snapshot verification for this change, one must
(1) discover all traffic classes on link A, (2) create a specification asserting that the discovered traffic classes traverse link B in the
new network, 
(3) discover \emph{all other traffic classes} and \emph{all their current paths}, exactly, (4) 
create a specification asserting all such other traffic classes continue to follow these discovered paths.  
%
\OMIT{
However, "single-snapshot verification" is not effective at precisely reasoning about network changes because it needs a detailed spec of the post-change snapshot's intended behavior, but such specs are rarely available. Consider a common network change that moves all traffic traversing link A to link B as a precursor to shutting off A for maintenance. The engineer wants to ensure that all traffic traversing link A is moved, it is moved to link B and nowhere else, and that no other traffic is impacted. Writing this spec for single-snapshot verification requires that the engineer have precise information about all traffic that traverses link A and of all traffic that does not (to ensure that it is not impacted). Worse, they also need to know precise paths for all traffic classes (sets of flows that take the same path through a network) that traverse link A to ensure that no other aspect of their path, before link A and after link B, changes. 
}

Creating such specifications is almost impossible for most real-world networks~\cite{batfish-2023}. One challenge is \emph{scale}:  
The specification needed scales with the size of the network, and, of course, modern networks are enormous and continue to grow.  The network in our experiments has on the order of $10^3$ routers, $10^4$ routes per router, and $10^\flowscale$ classes of flows with distinct forwarding paths, with up to $10^4$ classes impacted by typical changes.  Making matters worse, there is an additional challenge of \emph{incomplete information}:
Networks evolve incrementally over years, and their size and complexity demand that different parts be managed by different teams; any given engineer will have only partial knowledge of a network's behavior. 
Creating a detailed specification in these circumstances and maintaining it through successive changes would require otherworldly effort.
%
\OMIT{
Unfortunately, many large, real-world networks suffer from two problems that make creating such specifications is next to 
impossible:
\textit{(1)~Incomplete information:} The networks are evolve incrementally over years and their size and complexity demands that different parts be managed by different teams. Consequently, any engineer tasked with implementing a change will have only partial knowledge of a network's behavior. 
\textit{(2)~Scale:} The networks are large and continue to grow. The network in our experiments has on the order of $10^3$ routers, $10^4$ routes per router, and $10^7$ traffic classes, with $10^5$ classes impacted by typical changes. Even if one creates a detailed spec for this network through some out-worldly effort, maintaining it through successive changes is even harder.
}

The upshot is that while single-snapshot verification helps ensure coarse, long-term invariants, 
it is not helpful when it comes to the fine details of many network updates. Yet network engineers must check such details because their violations create congestion-induced outages, security problems, or performance issues.
%
Lacking appropriate tools, network engineers today rely on manually inspecting the impact of changes. Unsurprisingly, manual inspection is time-consuming, tedious, and error-prone, sometimes taking many weeks and multiple attempts to check even simple-seeming changes. See section \Cref{sec:motivation} for a representative example. 

We introduce {\em relational network verification} 
and investigate whether it can make network changes more reliable, more efficient, and less dependent on
manual audits. 
Rather than reasoning about the behavior of a \emph{single} snapshot in isolation, relational network verification reasons about the similarities and differences (i.e., the \emph{relationships}) 
between the behavior of \emph{two} network snapshots.  

Relational specifications make it easy to specify "no change" for the traffic and paths engineers do not want to modify (and may not even know about).  Indeed, the size of a relational
network specification is proportional to the complexity of the change rather than that of the network as a whole.  If a desired network change is small (e.g., changing link A to link B), the
relational specification will also be small.  
It is no wonder then that engineers already informally use
such ideas to communicate their intent 
in change request tickets.  In a sense, relational specifications
capture formally the kind of thinking that engineers use, and in a way that allows automatic checking.
%
\OMIT{
It can better handle the twin challenges above. Detailed information about the behavior of the post-change network snapshot is not needed if the engineers can simply specify "no change" for traffic and (sub) paths they do not want to modify or do not know about. Such a specification is fundamentally a statement about two network snapshots and cannot be supported by single-snapshot verification. Further, for traffic and paths that do change, the impact of the change can be compactly described, even if the full network is huge, because any given change is only concerns a small subset of the network. This small impact is what makes it possible to communicate the intent of the change (e.g., in a change request ticket).
}

Realizing relational network verification requires (1) a formal specification language for compactly describing the intent of a change, and (2) a decision procedure to verify that the pre- and post-change network snapshots adhere to the specification. We develop a tool called \sysname with these capabilities. Network engineers use regular expressions to represent paths and simple modifiers such as adding or removing parts of the path to specify change intent. \sysname compiles this user-friendly language to a low-level, regular intermediate representation (RIR) based on the theory of regular languages and relations~\cite{hopcroft-ullman}. 
Our tool combines the generated RIR with data from the pre- and post-change network snapshots, and checks that the pair of snapshots satisfies the RIR specification by reducing the problem to equivalence-checking for finite state automata.
The final result is either a "thumbs up" (if the network satisfies the specification) or a set of counterexample flows and paths
(otherwise).

We evaluate \sysname using data from all complex, high-risk changes to a global backbone network over the last seven months. We find that \sysname specs are compact; 93\% of the changes need fewer than 10 terms in the language. And even though the network has over $10^3$ routers, validation take under 20 minutes for 80\% of the changes. 
We are now integrating \sysname into the change pipeline of this network.


\sysname barely scratches the surface when it comes to realizing the potential of relational network verification. It focuses on dataplane verification for networks with stateless forwarding, the same context that was targeted by the first wave of single-snapshot verification tools~\cite{stateless-netverify}. We expect that relational verification will prove effective in other contexts as well, including stateful forwarding and control planes. We also expect that it can help verify if two parts of the same snapshot are similar (e.g., two geographic regions), modulo a few exceptions~\cite{selfstarter}. Beyond verification, the compact change specs of \sysname may also enable automatic synthesis of network changes.  We look forward to exploring these topics in the future. 


\smallskip
\noindent {\bf Ethics:}{\em This work does not raise any ethical issues.}

\section{Network Changes Today}
\label{sec:motivation}

\begin{figure*}
    \begin{subfigure}{.24\textwidth}  
        \centering
        \includegraphics[width=.9\textwidth]{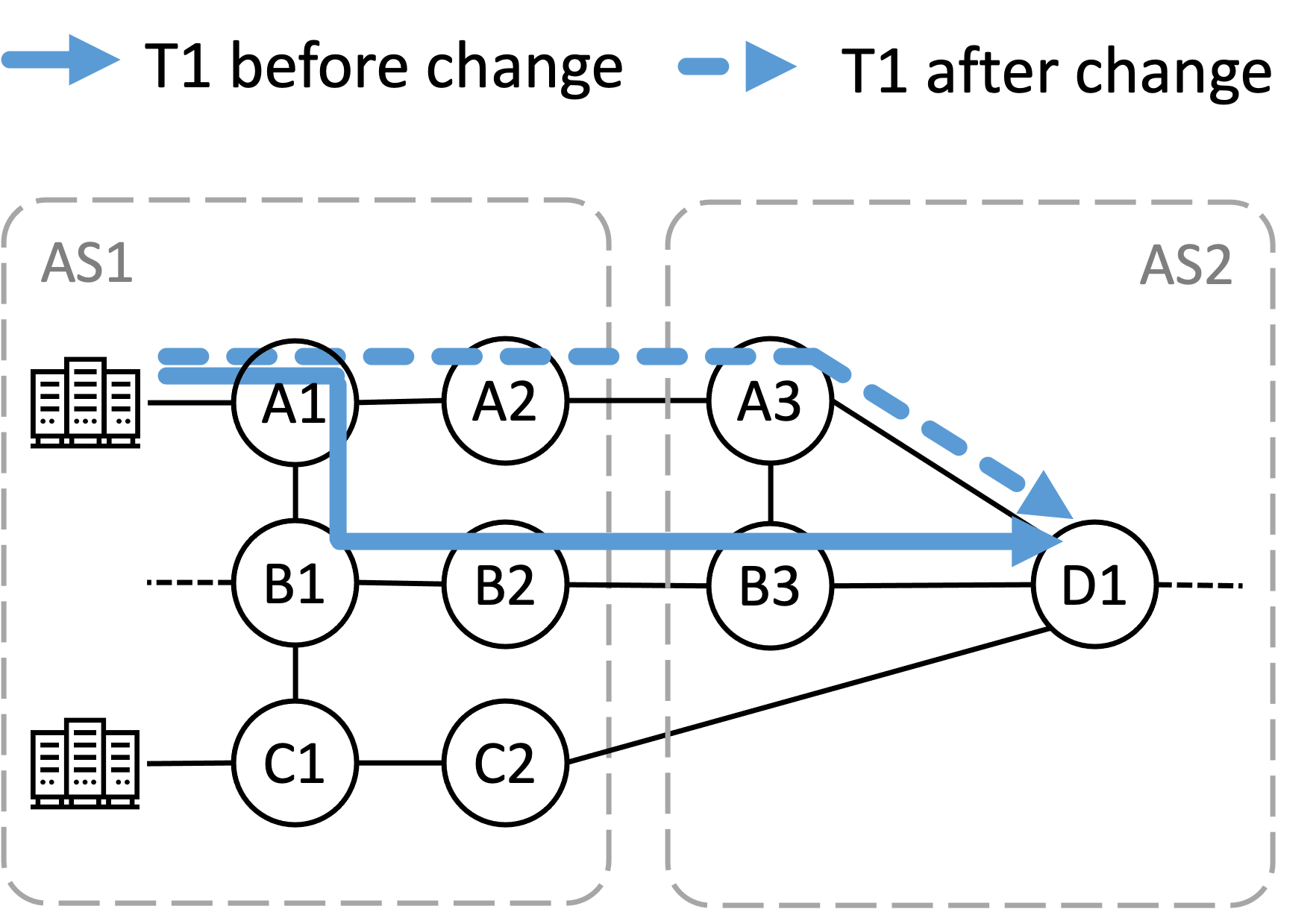}
        \caption{Change intent}
        \label{fig:shift_exit_intended}
    \end{subfigure}
    \begin{subfigure}{.24\textwidth}  
        \centering
        \includegraphics[width=.9\textwidth]{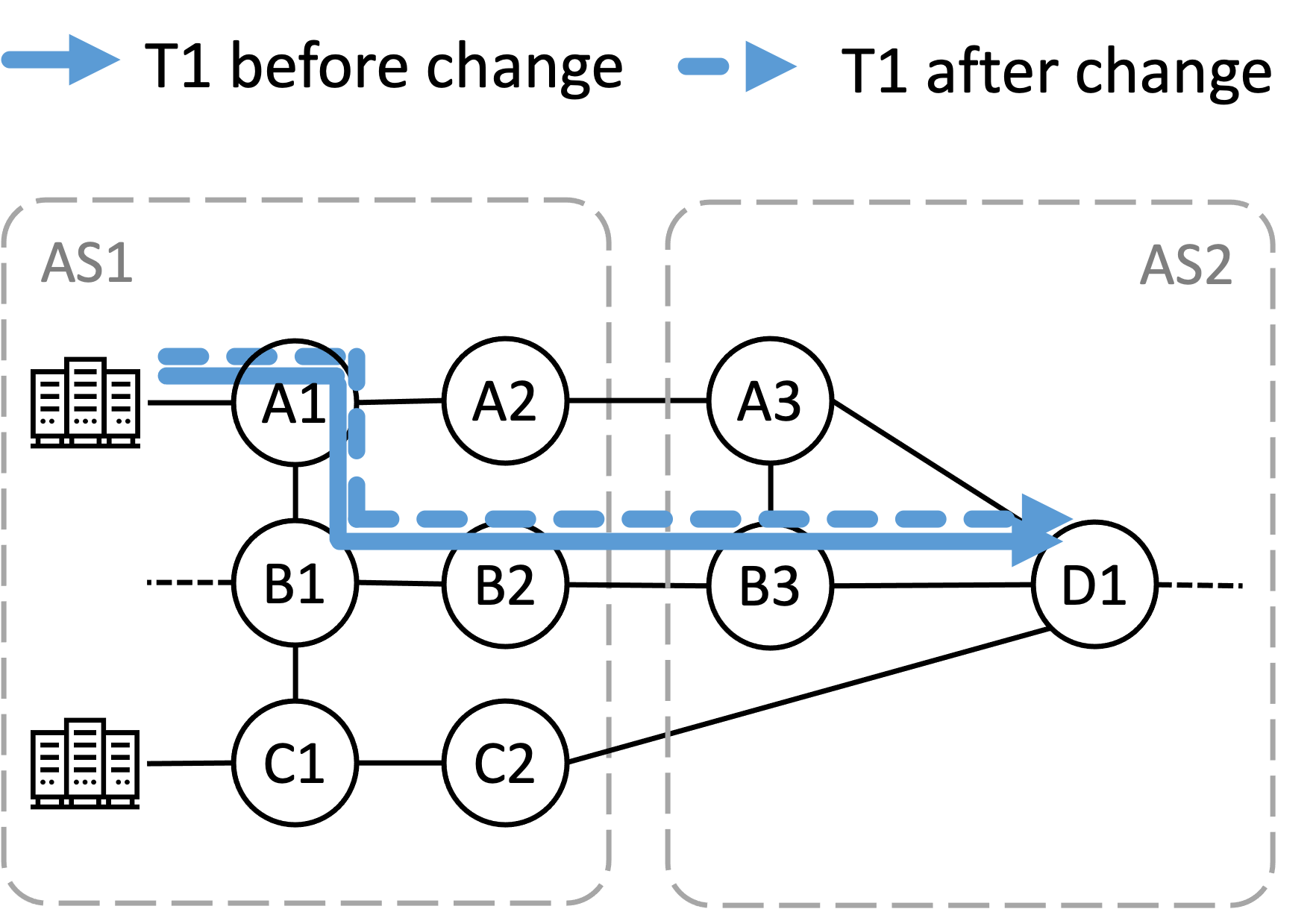}
        \caption{v1: Unchanged path}
        \label{fig:shift_exit_v1}
    \end{subfigure}
    \begin{subfigure}{.24\textwidth}  
        \centering
        \includegraphics[width=.9\textwidth]{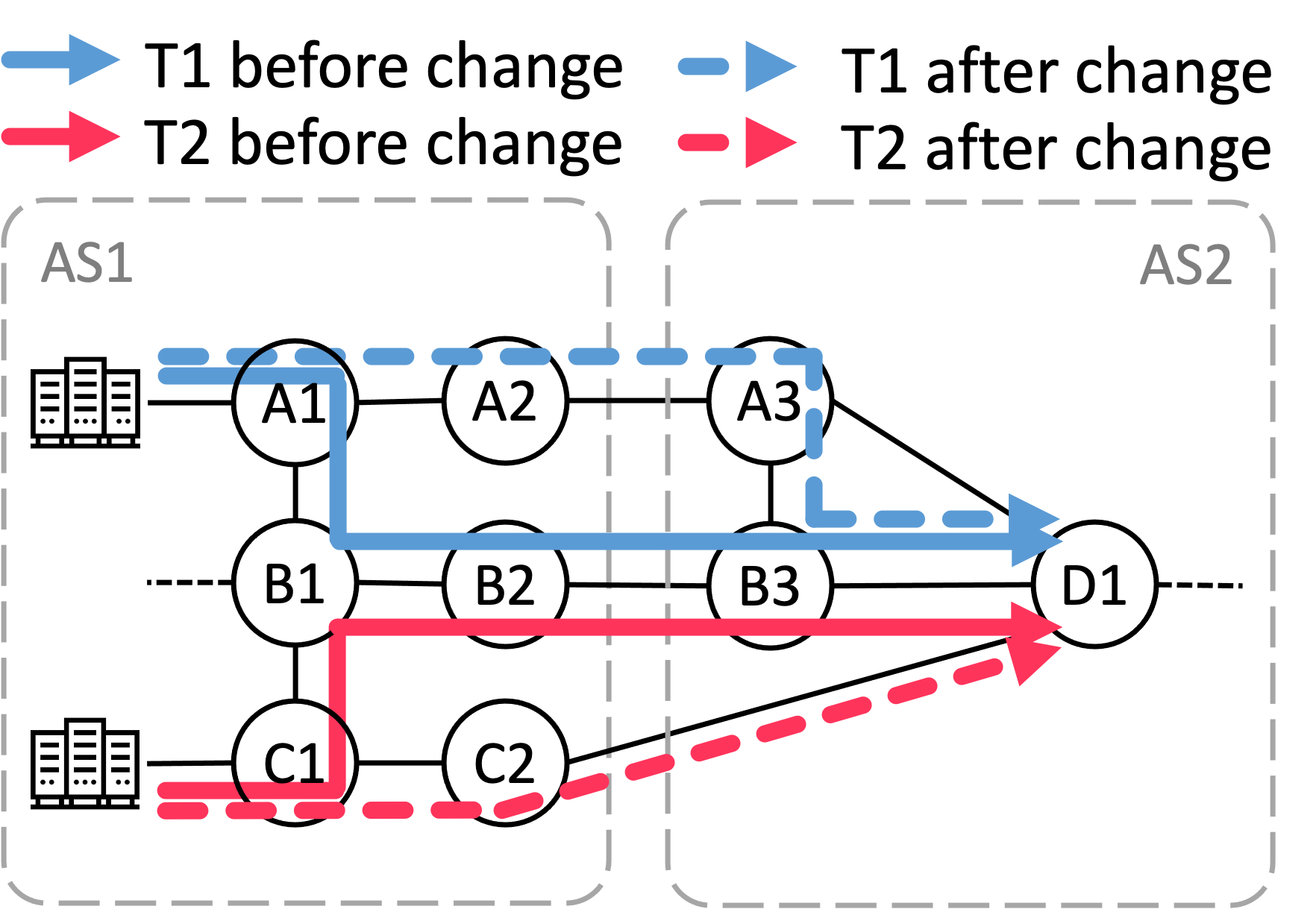}
        \caption{v2: Collateral damage}
        \label{fig:shift_exit_v2}
    \end{subfigure}
    \begin{subfigure}{.24\textwidth}  
        \centering
        \includegraphics[width=.9\textwidth]{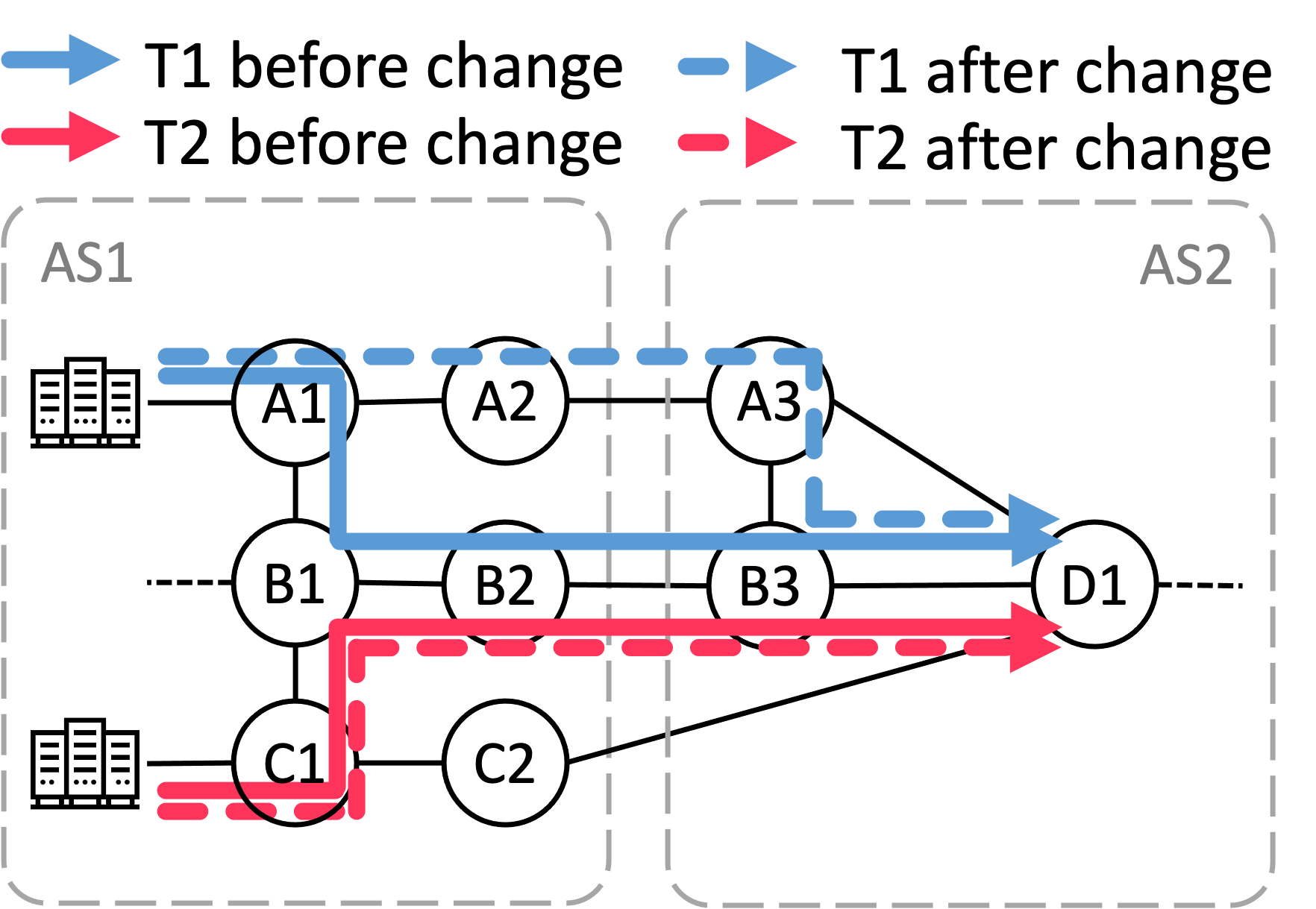}
        \caption{v3: Still incorrect path}
        \label{fig:shift_exit_v3}
    \end{subfigure}
    \caption{An example network change in a global WAN. T1 and T2 denote aggregate traffic bundles.}
    \label{fig:shift_exit}
\end{figure*}

Implementing network changes requires that engineers translate their network-wide intents into specific device-level configuration changes. Unfortunately, errors in translation between
high-level intent and low-level implementation are common. Using a change from a large
cloud provider's backbone, we illustrate the difficulty of making even seemingly simple changes and how incomplete information and scale limit the effectiveness of existing network analysis tools.

%


\subsection{An Example Change}
\label{sec:example}

\Cref{fig:shift_exit_intended} shows a change in the global backbone of a large cloud provider:
The blue (solid) line denotes a path in the old network, and the orange (dotted) line denotes the new path desired for that traffic.
Despite the simplicity of this abstract picture, it took network engineers \emph{four iterations across three weeks} to devise a working implementation of the change. 

The part of the backbone shown here has two BGP autonomous systems, $AS1$ and $AS2$, each enclosed by a grey box and had many routers. Each circle denotes a group of routers that fulfill the same functionality.  An AS spans multiple geographic regions, encoded using the prefix letter of router groups. So, $A1$ and $A2$ are in the same region, which is different from that of $B1$ and $B2$.  

The goal of the change is to prevent traffic, denoted $T1$, from region A from traversing region B while on its way to region D.
In order to do so, all traffic on the path $A1$-$B1$-$B2$- $B3$-$D1$ should move to $A1$-$A2$-$A3$-$D1$.  
Importantly, though the picture does not specify this intent overtly, no other WAN traffic should be impacted.  

\para{First iteration}
The engineers' first iteration (\Cref{fig:shift_exit_v1}) changed the configuration of $A2$ routers. 
They added $T1$ prefixes to an allow-list on $A2$, with the hope that $A1$ would pick the shorter path $A1$-$A2$ over $A1$-$B1$-$B2$.  However, on inspecting the impact of the change (using the process in \Cref{sec:strawman_manual}), the engineers found it ineffective: The $T1$ traffic 
followed the same path as before!
Investigation revealed that the routers in region $B$ were configured to announce T1 prefixes with a 
high local preference.  Since local preference overrides path length in BGP, $A1$ continued to prefer the route through $B1$ over $A2$.  This failure illustrates the challenge of \emph{incomplete information}: The engineers for region $A$ lack knowledge of how region $B$ routers are configured.

\para{Second iteration}
The engineers' second iteration (\Cref{fig:shift_exit_v2}) reconfigured $A2$ to increase the local preference of $T1$ prefixes. As a fail-safe, they also configured routers in region $B$ to lower the local preference for these prefixes. This time, the engineers observed that $T1$ had indeed moved from $B2$ to $A2$. However, it turned out that the implementation caused \emph{collateral damage}: the path of traffic T2, which should not have been impacted, changed. 
Debugging revealed that the root cause was a typo in the import policy at $B2$. 

\para{Third iteration}
The next iteration (\Cref{fig:shift_exit_v3}) fixed the typo. Upon testing, the engineers saw that it fixed the collateral damage but found another issue. While T1 traffic had indeed moved away from $B2$, it was bouncing back to $B3$ (due to an old configuration bug that made the link costs of $A3-B3-D1$ lower than those of $A3-D1$). It turns out that this undesirable behavior was present in the last implementation as well, but the engineers missed it amidst the information overload created by the collateral damage. 
This failure illustrates the challenge of \emph{scale}: It is not enough to focus on one small set of paths because even small configuration changes can impact many paths at once.

\para{Fourth iteration}
The fourth iteration finally achieved the intended behavior, after three weeks of labor. 

\smallskip
\noindent
Changes like this one are common in the backbone's daily operation. While some errors are caught prior to deployment, others make it to the network and have widespread impact. 

\subsection{Just Verify It?}
\label{sec:strawman_solutions}

Readers familiar with network verification might ask: Do the backbone network engineers have access to a verification tool; and does it help find errors in changes? The answers are: Yes, they have a verification tool, and they use it for certain tasks; and no, it does not help uncover the types of errors above. We explain why.

Abstractly, existing network verification methods operate as follows:  Given a specification $S$, check whether a network configuration $C$ satisfies $S$.  We call this method \emph{single-snapshot verification}  
because it analyzes a single network configuration against the specification.
Typically, one analyzes the new (post-change) network configuration, and the original (pre-change) network configuration is not used. 

Single-snapshot verifiers are deployed to check coarse properties that hold over long periods of time, such as "never block DNS traffic" and "always block ssh from outside." These verifiers can validate such properties are not violated by a proposed change, even when
networks are very large. 
However, to catch finer-grained problems, such as the problems with specific paths, described in the previous subsection, much finer-grained specifications are needed.  Unfortunately, with $10^\flowscale$ traffic classes, creating
a detailed specification for all of them is an insurmountable barrier to even getting started with 
single-snapshot network verification.  Said differently, the cost of creating network-wide single-snapshot specifications is proportional to the size of the network.  To make verification worthwhile,
we need to create specifications at a lower cost, ideally proportional to the size of the change.

One naive tactic to address this challenge is to generate small but highly incomplete single-snapshot specifications.  For instance,
if a network engineer wishes to replace a path $P1$ with $P2$, they might verify that $P2$ exists in the new network and $P1$ does not. But this tactic omits a key property: all other traffic should remain unchanged, and hence does not help identify any collateral damage that may have occurred.

For these reasons, while single-snapshot verification has a role in ensuring network reliability, it is insufficient for change validation. Our backbone's network engineers thus resort to other methods, which we discuss next. 

\subsection{Back to Manual Inspection :-( }
\label{sec:strawman_manual}


The dominant change validation method that engineers use today is manual inspection. Its workflow is: 
\begin{enumerate}
    \item use a simulator~\cite{hoyan,batfish} to compute the network's forwarding state $N1$, based on the current configuration;
    \item compute the network's after-change forwarding state $N2$, based on planned changes to the configuration;
    \item use $N1$ and $N2$ to compute the before- and after-change forwarding paths for all flows that traversed the network over the last hour;\footnote{NetFlow~\cite{netflow} monitoring provides this data. Engineers prefer it over considering all possible flows (i.e., symbolic analysis) because it reduces information they need to inspect and helps focus on flows that matter.} a \textit{flow} is a 5-tuple that starts at a particular point in the network;  
    \item aggregate flows into {\em equivalence classes} that contain flows with identical paths in each before- and after-change configuration; 
    \item manually inspect the {\em path diff}, which contains all equivalence classes whose paths differ for the two configurations, and check that all expected changes have occurred and no unexpected changes have occurred.  
\end{enumerate}
%
%
%
Manual auditing is a tedious, mind-numbing affair subject to human error.  Of course, the difficulty of conducting
an audit depends in large part on the size of the path diff.  Unfortunately, the
path diffs can vary anywhere from tens of differences to over 10,000.
Experienced engineers can audit only tens of classes per day, which makes a complete audit intractable for some changes.
Engineers may thus have to resort to sampling, increasing the risk of missing problems. 
Further, while it is relatively easy (but still hard) to ensure that no undesired path changes occur by inspecting the path diff, ensuring that all desired path changes occur is harder. Spotting an omission from a path diff is more difficult than missing the presence of a bad change.

\section{A New Approach: \\ Relational Verification}
\label{sec:relational}



Relational network verification is inspired by today's manual approach and relational program verification research from the formal methods community (see Barthe~\cite{relational-verification} for an introduction).
Relational methods reason about similarities and differences in \emph{two} versions of a system, rather than considering one version in isolation.  Because changes to a network involve
two network configurations, one old and one new, these methods naturally apply.  



Relational verification is better suited to validating network changes than single-snapshot verification because it is 
relatively easy to construct precise yet compact specifications for changes, even in enormous networks.  Abstractly, to replace a path $P_1$ with $P_2$, a relational specification will declare that traffic flowing over $P_1$ in the old network
should flow over $P_2$ in the new network and that all other traffic should follow the same path(s) in both networks.  Such a specification takes just a few lines of code because relationally specifying "no change" (i.e., old equals new)
is trivial.  Indeed, the specification for the example change in the previous section is roughly 
10 lines of code even though it moved over $10^5$ flows.  
Importantly, "no change" specifications are inherently relational---they
make direct use of comparisons between old and new---and there is no single-snapshot analog.



\section{\sysname by Example}
\label{sec:overview}



\sysname has a new specification language to describe the relationship between the forwarding behavior of two network snapshots.
This section introduces the language using the change in \Cref{fig:shift_exit}. The next section formalizes its syntax and semantics. 

Recall that the intent of the change in \Cref{fig:shift_exit} has three elements: (1) only impact the traffic from region $A$ to $D$ that traverses $A1$ and $D1$; (2) change the forwarding sub-paths of this traffic from $A1$-$B1$-$B2$-$B3$-$D1$ to $A1$-$A2$-$A3$-$D1$, while leaving unchanged the sub-paths before $A1$ and after $D1$ (which may be unknown to the engineer making the change); (3) no other traffic should be impacted. 

\para{Change \Zone{s}} 
In \sysname, the first step in defining a change intent is to define the \emph{change \zone}. Informally, 
%
change \zone{s} allow users to create a focus area for the impact of a change and ignore behaviors outside of that focus.
Users define change zones using \emph{path patterns}, which are regular expressions over network locations. 
\OMIT{
\dpw{this next sentence doesn't work for me.  It seems to be arguing against some alternative
form of domain specification but doesn't explicitly say what that alternative is (perhaps a set of traffic classes?).  I would delete the sentence.
However, instead of deleting the sentence one might explicitly state the alternative sort of domain specification 
one has in mind and to argue why it is not good.  However, doing so will interrupt the description of what we have
done.  Such an alternative would be better described in some later discussion section.
}
Despite the large number of traffic classes that may be touched by a change, we have found that such path expressions
usually suffice to characterize the domain effectively.
}



A \emph{network location} identifies one hop in a forwarding path. In \sysname, forwarding paths and locations can be viewed at different levels of granularity, including at the interface level, the router level or the router group level. Users may choose the level of granularity that suits their needs.
Our example uses router-level locations; our user does not care which interfaces are used for forwarding as long as they belong to the correct router.

\sysname is used in concert with a database that stores information about all locations available in the network.
Users can refer to a set of locations within the same entity (such as a router group or a tier)
by issuing "where" queries to select locations from the database and return the union of them.
We define below \lstinline{a1} to be the set of
routers with group attribute $A1$.
A similar query defines \lstinline{d1}.
\begin{lstlisting}
regex a1 := where(group=="A1")
regex d1 := where(group=="D1")
\end{lstlisting}
Regular expressions \lstinline{a1} and \lstinline{d1} can now be used to refer to routers in $A1$ and $D1$ in the rest of the \sysname specification.
For instance, the regex \lstinline{a1.*d1} denotes the set of paths that starts from any location in $A1$ and ends at any location in $D1$ after traversing zero or more (any) intermediate locations.


\OMIT{
The regular expression \lstinline{A1.*E1} expresses paths that start at $A1$, traverse any sequence of locations ($.^*$), and end at $E1$. One problem is that $A1$ or $E1$ is not a single (router) location, but many in a router group. \sysname provides \lstinline{where()} function to help specify a set of locations, as follows:
\begin{lstlisting}
regex a1 := where(group=="A1")
regex e1 := where(group=="E1")
\end{lstlisting}
The \lstinline{where()} function takes SQL-like filters as argument and it yields locations from a database based on the given filters and then return the set of all matching locations. Therefore, regex \lstinline{a1.*e1} denotes the set of paths that starts from any location in $A1$ and ends at any location in $E1$.
}

\para{Change specifications}  An atomic \emph{change specification} is written \textit{\zone} : \textit{modifier}. 
Roughly speaking, such a specification indicates that paths in the \zone should be changed according to the modifier.
When desired, such specifications may be named and reused or composed with other change specifications.  For instance:
\begin{lstlisting}
spec name := {zone : modifer}
\end{lstlisting}

Path modifiers describe the sets of paths to add, remove, replace, or preserve
between old and new network snapshots.  
For example, the following code presents one implementation of the second element of our example change intent.
\begin{lstlisting}
spec pathRepl := {
    a1.*d1 : replace(a1b1b2b3d1, 
                     a1a2a3d1)
}
\end{lstlisting}
The spec states that
for each path in the old network appearing in the \zone that matches \lstinline{a1b1b2b3d1}, all paths in \lstinline{a1a2a3d1} 
should appear in the new network (assuming symbols \lstinline{a2}, \lstinline{b1}, etc., have all been defined earlier as the union of routers in the corresponding router group).  
The semantics of "replace" also demands that if \lstinline{a1a2a3d1} appears in
the old network, it continues to appear in the new network.  

The replace modifier demands \emph{all} paths in \lstinline{a1a2a3d1} appear in the new network snapshot
if any path in \lstinline{a1b1b2b3d1} appears.
This may be what the user wants in some cases, but it may not be in others.  After all, \lstinline{a1a2a3d1} represents the Cartesian product of four router groups and contains a large number of possible paths---does the user want all such paths to be present in the new
network?  
The initial informal English specification we gave is actually
mute on this issue; it simply says "change it."  
Indeed, we have found that working with \sysname requires we think very carefully about exactly what we require, and typically, there are many corner cases to consider.  Still, because the specifications are so short (as well as reuseable and re-executable), one
can afford to think carefully about their consequences.

Fortunately, \sysname provides several different built-in modifiers if "replace" is not the desired one.
If the traffic should move to \emph{some} path ("any" of them) in \lstinline{a1a2a3d1}, an engineer can use the \lstinline{any(regex1)} modifier, as follows.
\begin{lstlisting}
spec pathShift := {
    a1.*d1 : any(a1a2a3d1);
}
\end{lstlisting}

Recall that traffic in our change zone may start upstream of $A1$ routers and continue downstream of $D1$ routers. The spec above has not expressed changes expected for these starting and ending \textit{sub-paths}. The user
may not even know all the paths leading to this part of the network.  In other systems, specifying a change accurately
 with such incomplete information is challenging, or perhaps impossible. Fortunately, though,
\sysname is \emph{compositional} as well as relational: One may stitch together change specifications of different kinds for
different subpaths to construct an end-to-end specification.  In this case, to specify that the beginnings and ends of our paths
should not change, we can use change specifications with the \lstinline{preserve} modifier as follows.
%
\begin{lstlisting}
spec e2e := {
    a* : preserve;
    pathShift;
    d* : preserve;
}
\end{lstlisting}
This spec, which concatenates three sub-path specs, defines the change zone as 
"\lstinline{a* (a1.*d1)  d*}". 
The first sub-spec's zone is \lstinline{a*}, which denotes arbitrary length paths within region $A$. Even though users may not know the details of sub-paths in this zone, they do understand that these sub-paths are expected to remain unchanged, and the \lstinline{preserve} modifier does the trick. We then reuse \lstinline{pathShift} defined earlier to specify the sub-path in the middle. And the spec of the third and last sub-path is similar to the first one. \sysname thus allows a precise end-to-end spec to be expressed compositionally, even when some parts of the paths are unknown to the users. 

Up to this point, we have a spec that defines which paths should change and how they should change. Our third and final task is to specify that no other paths are affected by the network update. Once again, \sysname makes this task easy via composition of specs using the \lstinline{else} operator:
\begin{lstlisting}
spec nochange := { .* : preserve; }
spec change := e2e else nochange
\end{lstlisting}
All traffic that does not match the first spec will fall through to the next spec chained by \lstinline{else}. Thus, all existing traffic except those matched by \lstinline{e2e} will be required to comply with \lstinline{nochange}---it must stay the same.

\para{Summary}
\sysname specifications describe relations between a pair of network snapshots---that is, the paths that are added, removed,
replaced or preserved when an old network is updated. It allows change zones to be defined at a level of location granularity appropriate to their task. Once a zone of interest is defined, one may craft atomic change specifications that
describe the relation between old and
new networks for (sub-)paths in a zone.  Users may draw on a collection of pre-defined modifiers to define relations of interest.  Finally, complex
change specs may be built out of simple ones through the use of \sysname's composition operators.

\section{Formalizing \sysname specifications}
\label{sec:formal}

This section specifies the formal syntax of \sysname and  
provides its semantics via translation
to an intermediate representation with \emph{regular relations}, which we
call the RIR.  While the RIR is more expressive than 
\sysname's surface language, it is also lower-level, making it harder to use by network engineers.  Indeed, \sysname was created with a goal of
making it easier to write relational specifications for networking
use cases.  Still, an
expert user may use the RIR directly if they choose.

\OMIT{
This section formalizes \sysname. The syntax in \Cref{sec:overview} uses \sysname's front-end (FE) language which is oriented toward ease of use by network engineers. We compile it to an regular intermediate representation (RIR) based on the theory of regular languages and regular relations \todo{cite}. 
The RIR design targets expressiveness and facilitates the development of a general and performant decision procedure (\Cref{sec:verification}). 
We now specify the two languages and the compilation strategy. 

\sysname targets networks that do not modify packet headers (e.g., NATs) or have stateful forwarding (e.g., firewalls). This is typical of backbones and interior fabrics of datacenters. In these networks, the primary concern is traffic paths, and thus the core language focuses on path changes. We will extend \sysname to other types of networks in the future.
}

\subsection{\sysname Syntax}\label{sec:language_fe}
\begin{figure}
  \centering
  \begin{bnf}(comment = {\#}, new-line-delim = {\|\|})[
    colspec={llcl@{}l},
    column{4}={font = \ttfamily},
  ]
    $r$ # Path Sets ::=
    || $l$               # location
    || ($r_1$|$r_2$)     # union
    || $r_1$$r_2$        # concatenation
    || $r$*              # Kleene star
    ;;
    $m$ # Modifiers ::=
    || \lstinline{preserve}
    || \lstinline{add}($r$)
    || \lstinline{remove}($r$)
    || \lstinline{replace}($r_1$,$r_2$)
    || \lstinline{drop}
    || \lstinline{any}($r$)
    ;;
    $s$ # Specs ::=
    || $r$:$m$           # atomic spec
    || $s_1$$s_2$       # concatenation
    || $s_1$ \lstinline{else} $s_2$  # prioritized union
  \end{bnf}
  \caption{The syntax of \sysname's front-end language.} 
  \label{fig:syntax_fe}
\end{figure}

\Cref{fig:syntax_fe} presents the formal syntax of \sysname, which includes
sublanguages for (regular) sets of paths ($r$), modifiers ($m$) and
specifications ($s$). This syntax omits named definitions
\texttt{spec n := \{ s \}}, which are easily inlined.  It also
excludes {\textbf{\texttt{where}}} queries to select locations from database, which are implemented as a prepass.  

We saw several of the modifiers in the previous section.  One that we did not see
is \lstinline{drop}, which indicates a path that should drop a packet. We model this behavior as a special path with a single location "drop".
Each modifier is defined
by a straightforward translation into the RIR.  
While our
experiments suggest that we have developed a useful set of modifiers,
adding new ones is not difficult, provided that they can be encoded in
the RIR.  

\subsection{Regular IR (RIR)}\label{sec:language_rir}

\begin{figure*}
  \centering
  \begin{bnf}(comment = {\#}, new-line-delim = {\|\|})[
    colspec={llcll},
    column{4}={font = \ttfamily},
  ]
    $P \in Path\ Set$  ::=
    || $a ~|~ 0 ~|~ 1 ~|~ \mathtt{PreState} ~|~ \mathtt{PostState}
    ~|~ (P_1 | P_2) ~|~ P_1P_2 ~|~ P^* ~|~ P_1 \cap P_2 ~|~ \overline{P}  ~|~ P \triangleright R$
    ;;
    $R \in Rel$ ::=
    || $P_1 \times P_2 ~|~ \mathtt{I}(P) ~|~ 0 ~|~ 1  ~|~ (R_1 | R_2) ~|~ R_1R_2 ~|~ R^* ~|~ R_1 \circ R_2  $
    ;;
    $S \in Spec$ ::=
    || $P_1 = P_2 ~|~ P_1 \subseteq P_2 ~|~ S_1 \vee S_2 ~|~ S_1 \wedge S_2 ~|~ \neg S$
  \end{bnf}
\end{figure*}
\begin{figure*}
    \begin{minipage}[c]{0.4\textwidth}
        \begin{align*}
        \multicolumn{2}{l}{\fbox{Path Sets}} \\
\mathscr{P}\llbracket a \rrbracket (M, N) &\triangleq \{a\} \\
\mathscr{P}\llbracket 0 \rrbracket (M, N) &\triangleq \emptyset \\
\mathscr{P}\llbracket 1 \rrbracket (M, N) &\triangleq \{\epsilon\} \\
\mathscr{P}\llbracket \mathtt{PreState} \rrbracket (M, N) &\triangleq M \\
\mathscr{P}\llbracket \mathtt{PostState} \rrbracket (M, N) &\triangleq N \\
\cdots & \\ 
\mathscr{P}\llbracket P \triangleright R \rrbracket (M, N) &\triangleq \{q:\exists p.~\langle p,q \rangle \in \mathscr{R}\llbracket R \rrbracket (M, N) \\
& \qquad\qquad\qquad \wedge p \in \mathscr{P}\llbracket P \rrbracket (M, N) \} \\
\end{align*}
    \end{minipage}\quad\begin{minipage}[c]{0.4\textwidth}
\begin{align*}
\multicolumn{2}{l}{\fbox{Relations}} \\
\mathscr{R}\llbracket P_1 \times P_2 \rrbracket (M, N) &\triangleq \{\langle p_1, p_2 \rangle \mid p_1 \in \mathscr{P}\llbracket P_1\rrbracket (M, N),\\
& \qquad\qquad\ \ \ \ \ \  p_2 \in \mathscr{P}\llbracket P_2\rrbracket (M, N)\} \\
\mathscr{R}\llbracket \mathtt{I}(P) \rrbracket (M, N) &\triangleq \{\langle p, p \rangle \mid p \in \mathscr{P}\llbracket P\rrbracket (M, N)\}\\
\mathscr{R}\llbracket 0 \rrbracket (M, N) &\triangleq \emptyset \\
\mathscr{R}\llbracket 1 \rrbracket (M, N) &\triangleq \{(\epsilon,\epsilon)\} \\
\mathscr{R}\llbracket R_1 | R_2 \rrbracket (M, N) &\triangleq\mathscr{R}\llbracket R_1\rrbracket (M, N) \cup \mathscr{R}\llbracket R_2 \rrbracket (M, N) \\
\cdots \\
\multicolumn{2}{l}{\fbox{Specifications}} & \\
M, N \models P_1 = P_2 &\Longleftrightarrow \mathscr{P}\llbracket P_1 \rrbracket (M, N) = \mathscr{P}\llbracket P_2 \rrbracket (M, N) \\
M, N \models P_1 \subseteq P_2 &\Longleftrightarrow \mathscr{P}\llbracket P_1 \rrbracket (M, N) \subseteq \mathscr{P}\llbracket P_2 \rrbracket (M, N) \\
\cdots \\
\end{align*}
    \end{minipage}
    \vspace{-20pt}
  \caption{RIR Syntax (top) and semantics of selected features (bottom).}
  \label{fig:semantics_rir}
\end{figure*}



The \sysname RIR is an intermediate language
for defining \emph{regular sets} of paths and \emph{regular relations} between 
paths.  A regular set is a set created through the usual 
operations on regular languages (concatenation, union, and Kleene star).
Likewise, regular relations are binary relations between paths
(i.e., sets of pairs of paths), also
constructed with the usual operations on regular languages. 
Since all RIR-expressible sets and relations are regular, we are able to make use of known, efficient constructions and decision procedures from automata theory as the basis of a decision procedure for RIR.


\Cref{fig:semantics_rir}(top) presents the syntax of the RIR, which
contains three sublanguages.  The language of path sets ($P$) describes
regular sets of paths over the alphabet $\Sigma$, which includes the
set of network locations as well as the special "drop" symbol.  
We use $a$ to denote an arbitrary character from $\Sigma$.
%
The path sets $a$, $0$, and $1$ denote 
sets with a single one-hop path, no paths at all, and a single 
0-length path (written $\epsilon$). The special
symbol \lstinline{PreState} denotes the set of paths in the pre-change network.  Similarly, \lstinline{PostState} 
denotes the set of paths in the 
post-change network.\footnote{In principle, \lstinline{PreState}
and \lstinline{PostState} may refer to the set of \emph{all} paths
in the pre-change (post-change) networks respectively. In practice, to scale the Rela tool to networks
with $10^\flowscale$ traffic classes, we
apply the specification to every traffic class separately and
in parallel.}  The expressions $P_1|P_2$, $P_1P_2$, and $P^*$ denote union, concatenation, and Kleene star operations over path sets. Finally, $P \triangleright R$ denotes the {\em image}, the path set derived by applying relation $R$ to paths recognized by $P$.  
In other words, $P \triangleright R$ describes the set of paths that are related (via $R$) to \textit{some} path recognized $P$.
\Cref{fig:semantics_rir} (bottom left) presents selected equations defining the semantics of path sets.
These equations have the form 
$\mathscr{P}\llbracket P \rrbracket (M, N) \triangleq S$, meaning that 
$P$ describes the set of paths $S$ when $M$ is the pre-change network
snapshot and $N$ is the post-change network snapshot.

$Rel$ denotes regular relations, which are sets of pairs of paths. 
Alternately, a relation may be viewed as a map from a path to zero or more related paths.  The cross-product relation $P_1 \times P_2$ denotes
the relation that associates every path in $P_1$ with all paths in $P_2$.
The identity relation $\mathtt{I}(P)$ associates every path in $P$ with itself;
paths not in $P$ are not related to any other path by $\mathtt{I}(P)$.
The symbols $0$ and $1$ denote the empty relation and the relation
associating $\epsilon$ with itself.
$R_1|R_2$, $R_1R_2$, $R^*$, and $R_1\circ R_2$ denote union, concatenation, Kleene star, and composition of relations.  (Rational relations are closed 
under all of these operations~\cite{ElgotMezei1965}.) 
\Cref{fig:semantics_rir} (middle right) shows the semantics of relations.  The equations have
the form $\mathscr{R}\llbracket R \rrbracket (M, N) \triangleq T$,
meaning that 
$R$ describes set of pairs of paths $T$ when $M$ is the pre-change network snapshot and $N$ is the post-change network snapshot.



Finally, $Spec$ denotes specifications that relate sets of paths. 
Such specifications may include equality ($P_1 = P_2$), set inclusion
($P_1 \subseteq P_2$), and boolean combinations of such specifications. 
As an example, consider this spec: 
$$\mathtt{PreState} \triangleright R = \mathtt{PostState}$$ 
Assuming the relation $R$ is an intended transformation of the  network, the spec says that if one applies the transformation $R$
to the pre-change network, then one should obtain a result that equals
the post-change network.  Our translation from
\sysname's surface language into the RIR uses this sort of idiom.
\Cref{fig:semantics_rir} (bottom right) presents selected rules
from the semantics of
specifications.  Each rule has the form 
$M, N \models S \Longleftrightarrow Bool$, 
which may be read "pre-change snapshot $M$ and post-change snapshot $N$ satisfy $S$ if and only if $Bool$ is true."


The full semantics of RIR appears in \Cref{sec:semantics}.

\subsection{Compilation from \sysname to RIR}\label{sec:compilation}



To compile \sysname into the RIR, 
from each specification, we generate one relation to transform the pre-change network and another relation to transform the post-change network, and produce an equation that checks whether the results of those transformations are equal. More formally, the translation of a \sysname spec $s$ is an RIR expression of the following form.
$$\mathtt{PreState} \triangleright \mathcal{R}_{pre}\llbracket s\rrbracket = \mathtt{PostState} \triangleright \mathcal{R}_{post}\llbracket s\rrbracket $$
In what follows, we show how to compute relations for some of the key modifiers in the \sysname language.



\para{Encoding path preservation}
Consider the translation of the path preservation modifier ``\lstinline{D: preserve}''.
%
Intuitively, this change specification says that all paths that appear in the zone $D$ in the pre-state should also appear in the post-state. If the pre- and post-relations are as follows:
\begin{align*}
\mathcal{R}_{pre}\llbracket D: \mathtt{preserve} \rrbracket \triangleq \mathtt{I}(D)\\
\mathcal{R}_{post}\llbracket D: \mathtt{preserve} \rrbracket \triangleq \mathtt{I}(D)
\end{align*}
then our overall translation will be:
$$\mathtt{PreState}  \triangleright \mathtt{I}(D) = \mathtt{PostState} \triangleright \mathtt{I}(D) $$
which is equivalent to the equation:
$$(\mathtt{PreState} \cap D ) = (\mathtt{PostState} \cap D) \ ,$$
as desired.


\para{Encoding path additions} Consider adding the paths $P$ 
when the pre-change network contains a path in $D$.\footnote{
The Rela surface language can not express addition of a path in $D$ when the pre-change
network contains no path in $D$.  Such ``unconditional'' path additions can be expressed in the RIR, however.  For instance, the equation $\mathtt{PostState} = \mathtt{PreState}|P$ expresses that exactly the set of paths recognized by $P$ are added to the network.}
Our goal now is to preserve all of the paths in the zone from the pre-state into the post-state. In other words, we would like to apply the identity relation $\mathtt{I}(D \mid P)$. In addition, we would like a relation that adds the path $P$. We can use the relation $D \times P$ to do so. Overall, our pre-relation is the combination of those two relations. Hence we generate the following equations.
\begin{align*}
\mathcal{R}_{pre}\llbracket D: \mathtt{add}(P) \rrbracket \triangleq&~ \mathtt{I}(D \mid P) \mid (D \times P)\\
\mathcal{R}_{post}\llbracket D: \mathtt{add}(P) \rrbracket \triangleq&~ \mathtt{I}(D \mid P)
\end{align*}


\para{Encoding path removals} Next, consider path removals using the modifier ``\lstinline{D: remove(P)}''. This modifier expresses that the paths in $D$ in the pre-state should be preserved in the post-state, except the paths in $P$ which should be removed.
Hence, our relations are as follows.
\begin{align*}
\mathcal{R}_{pre}\llbracket D: \mathtt{remove}(P) \rrbracket \triangleq&~ \mathtt{I}(D \setminus P)\\
\mathcal{R}_{post}\llbracket D: \mathtt{remove}(P) \rrbracket \triangleq&~ \mathtt{I}(D)
\end{align*}


\para{Encoding non-deterministic path replacement} The modifier ``\lstinline{D: any(P)}'' demands that (1) if there is any path in $D \mid P$ in the pre-state, there must be some path in $P$ in the post-state and (2) all paths in $D \mid P$ in the post-state must be in $P$. 
To encode this condition, we use a relation for the pre-state that replaces paths in $D \mid P$ with a symbol $\#$. Likewise, the relation for the post-state replaces all paths in $P$ with $\#$, while also retaining the paths in $D \setminus P$.  Since paths in $D\setminus P$ are \textit{not} retained in the pre-state relation, this relation encodes that there are no paths in $D\setminus P$ in the post-state network. Together, the two relations enforce the desired condition.
\begin{align*}
\mathcal{R}_{pre}\llbracket D: \mathtt{any}(P) \rrbracket \triangleq&~ (D \mid P) \times \#\\
\mathcal{R}_{post}\llbracket D: \mathtt{any}(P) \rrbracket \triangleq&~ (P \times \#) \mid \mathtt{I}(D \setminus P)
\end{align*}

\para{Encoding prioritized union} A prioritized union ``$s_1~\mathtt{else} ~s_2$'' should apply the change specification $s_1$ to $s_1$'s zone and $s_2$ to everything else in $s_2$'s zone. To achieve this specification in the RIR, we need to explicitly extract $s_1$'s zone. We do so with an auxiliary function $\mathcal{Z}\llbracket D: \mathtt{modifier} \rrbracket$. 
See \Cref{fig:compilation_rules} for the full definition of $\mathcal{Z}\llbracket \cdot \rrbracket$.

To translate ``$s_1~\mathtt{else} ~s_2$'', we first translate $s_1$, and then take the union with the translation of $s_2$ applied exclusively to the complement of the zone of $s_1$.

\para{Summary} See \Cref{fig:compilation_rules} for the complete translation.

\begin{figure}
\begin{align*}
    \mathcal{R}_{pre}\llbracket D: \mathtt{preserve} \rrbracket &\triangleq \mathtt{I}(D) \\
    \mathcal{R}_{pre}\llbracket D: \mathtt{add}(P) \rrbracket &\triangleq \mathtt{I}(D ~|~ P) ~|~ (D \times P) \\
    \mathcal{R}_{pre}\llbracket D: \mathtt{remove}(P) \rrbracket &\triangleq \mathtt{I}(D\setminus P) \\
    \mathcal{R}_{pre}\llbracket D: \mathtt{replace}(P_1, P_2) \rrbracket &\triangleq \mathtt{I}((D~|~P_2) \setminus P_1) \\
    &\ ~|~((D \cap P_1)\times P_2)\\
    \mathcal{R}_{pre}\llbracket D: \mathtt{drop} \rrbracket &\triangleq (D ~|~ drop) \times drop \\
    \mathcal{R}_{pre}\llbracket D: \mathtt{any}(P) \rrbracket &\triangleq (D ~|~ P) \times \# \\
    \mathcal{R}_{pre}\llbracket s_1s_2 \rrbracket &\triangleq  \mathcal{R}_{pre}\llbracket s_1 \rrbracket  ~\mathcal{R}_{pre}\llbracket s_2 \rrbracket\\
    \mathcal{R}_{pre}\llbracket s_1~\mathtt{else}~s_2 \rrbracket &\triangleq  \mathcal{R}_{pre}\llbracket s_1 \rrbracket \\
     &\ ~|~ \left(\mathtt{I}(\widebar{{\mathcal{Z}}\llbracket {s_1} \rrbracket}) \circ \mathcal{R}_{pre}\llbracket s_2 \rrbracket \right)
\end{align*}
\begin{align*}
    \mathcal{R}_{post}\llbracket D: \mathtt{preserve} \rrbracket &\triangleq \mathtt{I}(D) \\
    \mathcal{R}_{post}\llbracket D: \mathtt{add}(P) \rrbracket &\triangleq \mathtt{I}(D ~|~ P) \\
    \mathcal{R}_{post}\llbracket D: \mathtt{remove}(P) \rrbracket &\triangleq \mathtt{I}(D) \\
    \mathcal{R}_{post}\llbracket D: \mathtt{replace}(P_1, P_2) \rrbracket &\triangleq \mathtt{I}(D ~|~ P_2)\\
    \mathcal{R}_{post}\llbracket D: \mathtt{drop} \rrbracket &\triangleq \mathtt{I}(D ~|~ drop) \\
    \mathcal{R}_{post}\llbracket D: \mathtt{any}(P) \rrbracket &\triangleq (P \times \#) ~|~ \mathtt{I}(D \setminus P)\\
    \mathcal{R}_{post}\llbracket s_1s_2 \rrbracket &\triangleq  \mathcal{R}_{post}\llbracket s_1 \rrbracket  ~\mathcal{R}_{post}\llbracket s_2 \rrbracket\\
    \mathcal{R}_{post}\llbracket s_1~\mathtt{else}~s_2 \rrbracket &\triangleq  \mathcal{R}_{post}\llbracket s_1 \rrbracket \\
    &\ ~|~ \left(\mathtt{I}(\widebar{{\mathcal{Z}}\llbracket {s_1} \rrbracket}) \circ \mathcal{R}_{post}\llbracket s_2 \rrbracket \right)
\end{align*}
\begin{align*}
    \mathcal{Z}\llbracket D: \mathtt{preserve} \rrbracket &\triangleq D \\
    \mathcal{Z}\llbracket D: \mathtt{add}(P) \rrbracket &\triangleq D ~|~ P \\
    \mathcal{Z}\llbracket D: \mathtt{remove}(P) \rrbracket &\triangleq D \\
    \mathcal{Z}\llbracket D: \mathtt{replace}(P_1, P_2) \rrbracket &\triangleq D ~|~ P_2\\
    \mathcal{Z}\llbracket D: \mathtt{drop} \rrbracket &\triangleq D ~|~ drop \\
    \mathcal{Z}\llbracket D: \mathtt{any}(P) \rrbracket &\triangleq D ~|~ P\\
    \mathcal{Z}\llbracket s_1s_2 \rrbracket &\triangleq  \mathcal{Z}\llbracket s_1 \rrbracket  ~\mathcal{Z}\llbracket s_2 \rrbracket\\
    \mathcal{Z}\llbracket s_1~\mathtt{else}~s_2 \rrbracket &\triangleq  \mathcal{Z}\llbracket s_1 \rrbracket  ~|~  \mathcal{Z}\llbracket s_2 \rrbracket
\end{align*}
    \caption{\sysname to RIR translation.}
    \label{fig:compilation_rules}
\end{figure}


\section{Decision procedure}
\label{sec:verification}





Given an RIR spec and two sets of forwarding paths (represented using DAGs), corresponding to the before- and after-change network snapshots, the decision procedure determines whether the path set pair meets the spec. When the pair does not, it provides counterexamples
in the form of specific flows and paths that violate the spec.

\subsection{RIR to Finite-State Automaton (FSA)} 
The first step in validating an RIR specification involves constructing a finite-state automaton (FSA) from each $Prop$ and $Rel$ expression. 
Per Kleene's Theorem, every regular language can be represented by an FSA that moves from one state to another in response to the input sequence of symbols. Similarly, every regular relation can be represented as a finite-state transducer (FST) \cite{ElgotMezei1965}.

An FST is essentially an FSA that uses two tapes. It may be viewed as a ``translating machine'' that reads from an input tape and writes to the output tape. The following diagram presents an FST for relation  $a \times b$, which translates path $a$ into path $b$. 
$$
\begin{tikzpicture}[node distance = 1.5cm, on grid, auto]
\tikzstyle{every state}=[fill=white,draw=black,text=black,minimum size=20pt]
    \node (q0) [state, initial] {$q_0$};
    \node (q1) [state, accepting, right = of q0] {$q_1$};

    \path [->]
        (q0) edge node {a:b}   (q1);

\end{tikzpicture}
$$

\noindent
The label a:b on the arc means $a$ should be read from the input tape and $b$ should be written to the output tape. 


From small, simple FSAs, like the one above, we can build larger, more complex ones using standard automaton composition algorithms. 
%
In what follows, we sketch some of the algorithms used to construct \sysname-specific symbols and operators. Most other aspects
of the compilation strategy
are well-established, and are thus omitted
(see Thompson's construction \cite{Thompson1968}, for
instance).

\para{\texttt{PreState} and \texttt{PostState} symbols} Conceptually, the input to the decision procedure contains two sets of forwarding paths that corresponds to \texttt{PreState} and \texttt{PostState} respectively. In practice, however, the number of ECMP forwarding paths may explode in a large network. This problem is prominent when forwarding behavior is modeled at the interface-level, because the network may employ 10s of parallel links between any two hops to increase capacity. Indeed, we recorded a flow with $10^8$ interface-level ECMP paths for our backbone, and it takes several hours just to deserialize the paths from file input. To address this challenge, \sysname defines a graph format to represent the interface-level input path set. Each vertex in the graph denotes a router that appears as a forwarding hop for this traffic, and each directed edge denotes a physical link that is used to forward this traffic between the two hops. There is also extra metadata to identify all source vertices and sink vertices (start and end locations of paths) in the DAG. With this format, the $10^8$ paths of the aforementioned traffic class can be encoded with a DAG with 38 vertices and 50K edges. 

Constructing an FSA for \texttt{PreState} and \texttt{PostState} from the forwarding graph is straightforward: We turn vertices and edges in the DAG into FSA states and transitions respectively. If the user has specified a coarser granularity than interface-level (e.g., router level), we do granularity conversion in this step by merging vertices that belong to a same coarser entity. 
Next, we add a initial state $q_0$ and draw an $\epsilon$-transition from $q_0$ to each source node identified by the metadata. Finally, we set all sink nodes to be accepting states of the FSA.

\para{$P_1 \times P_2$ relation} 
The FST for $P_1 \times P_2$ may be obtained by
(1) translating the FSA for $P_1$ into an FST that accepts $P_1$ on its first tape and $\epsilon$ on its second, (2) translating the FSA for $P_2$ to a 
FST that accepts $P_2$ on its second tape and $\epsilon$ on its first, 
and (3) concatenating the results.  An illustration of this construction for  $a \times b$  can be found in figure below (recall that $\epsilon$ is the empty string).

$$
\begin{tikzpicture}[node distance = 1.5cm, on grid, auto]
\tikzstyle{every state}=[fill=white,draw=black,text=black,minimum size=20pt]
    \node (q0) [state, initial] {$q_0$};
    \node (q1) [state, right = of q0] {$q_1$};
    \node (q2) [state, accepting, right = of q1] {$q_2$};

    \path [->]
        (q0) edge node {$a:\epsilon$}   (q1);
    \path [->]
        (q1) edge node {$\epsilon:b$}   (q2);
\end{tikzpicture}
$$

\para{$\mathtt{I}(P)$ relation} The FST of $\mathtt{I}(P)$ is the same as the FSA of $P$ except that each FST transition has an output symbol that is the same as the input symbol.

\para{$P \triangleright R$ image} One may compile $P \triangleright R$ by translating it into the composition of two relations: $\mathtt{I}(P) \circ R$. The composition of two regular relations $R_1 \circ R_2$ 
may be compiled using standard FST algorithms~\cite{ElgotMezei1965}. 



\subsection{Compliance checking} 

Once we have the FSA representation of both sides of an equation, we can check for equality (or inclusion) using standard automaton equivalence algorithms. Once we have solved the basic equations (inequations), we can decide any boolean combination of them
by applying conjunction, disjunction, or negation operations as
specified.


\begin{table*}[t!]
    \centering
    \begin{tabular}{l|l|l|l}
    \toprule
        FEC & Pre-change paths & Post-change paths & Cause of violation\\ \midrule
        $\{(ip_1,~\mathrm{ingress}~= x_1)\}$   & $\{x_1A_1B_1B_2B_3D_1y_1\}$  & $\{x_1A_1A_2A_3B_3D_1y_1\}$ &  \texttt{e2e}: $\{x_1A_1A_2A_3D_1y_1\} \neq \{x_1A_1A_2A_3B_3D_1y_1\}$ \\ \midrule
        $\{(ip_2,~\mathrm{ingress}~= x_2)\}$ & $\{x_2C_1B_1B_2B_3D_1y_2\}$ & $\{x_2C_1C_2D_1y_2\}$ & \texttt{nochange}: $\{x_2C_1B_1B_2B_3D_1y_2\} \neq \{x_2C_1D_2D_1y_2\}$ \\
    \bottomrule
    \end{tabular}
    \vspace{5pt}
    \caption{A subset of counterexamples generated by \sysname when verifying the change implementation in \Cref{fig:shift_exit_v2}. The first row shows a flow in traffic class T1, and the second row shows a flow in T2.}
    \label{tab:counterexamples}
\end{table*}

\subsection{Counterexample Generation}

If a network change violates a \sysname{} specification, then
we generate an exhaustive list of counterexamples, where each entry is a flow equivalence class (FEC), the pre- and post-change paths for the FEC, and a reason that explains the failure. 
\Cref{tab:counterexamples} shows a subset of counterexamples reported by \sysname when verifying the change implementation in \Cref{fig:shift_exit_v2} using the \lstinline{change} spec in \Cref{sec:overview}. The two entries indicate incorrect path changes for traffic T1 and collateral damage for T2.

The forwarding paths that violate the specs are derived by extracting paths from the difference of two FSAs.
Recall that a \sysname{} spec $s$ is translated to an equation of the form $P_1 = P_2$ in RIR, where $P_1 = \mathtt{PreState} \triangleright \mathcal{R}_{pre}\llbracket s\rrbracket$ and $P_2 = \mathtt{PostState} \triangleright \mathcal{R}_{post}\llbracket s\rrbracket$.
The difference $P_1 \setminus P_2$ represents the expected forwarding paths that are missing from the observed post-change network, and $P_2 \setminus P_1$ represents the unexpected paths in the post-change network.
After extracting the violating paths, we find the flow to which each violating path belongs. We do so by extracting all paths with the same starting locations as the violating paths from \texttt{PreState} and \texttt{PostState}. \sysname aggregates all violating flows into equivalence classes in a manner similar to the manual inspection today (\Cref{sec:strawman_manual}) to aid analysis by engineers. 

For each violating flow, we generate a reason to help understand why it failed the spec. For specs that are composed using the \lstinline{else} operator, we can find the exact sub-spec that failed a flow by matching the flow with the zone of each sub-spec. We then apply $\mathcal{R}_{pre}$ and $\mathcal{R}_{post}$ of this sub-spec to the flow's pre- and post-change path set respectively. The difference of the two derived sets explains the failure of set equation and inclusion assertions made by the spec. For special symbols introduced by rewriting in the compilation process, we rewrite them back to their original forms to make the counterexamples more human-readable. For example, the before paths in the first row of \Cref{tab:counterexamples} yield $\{x_1\#y_1\}$ when applying $\mathcal{R}_{pre}$, where ``\#'' rewrites $A_1A_2A_3D_1$. After undoing this rewriting, the ``Reason of violation'' column clearly shows that the flow failed the sub-spec \lstinline{e2e}, which expected the path set to be $\{x_1A_1A_2A_3D_1y_1\}$ after change. This set is not equal to the observed path set $\{x_1A_1A_2A_3B_3D_1y_1\}$.



\section{Implementation}

We implemented \sysname with 6,000 lines of Python code. \sysname and RIR are implemented as domain-specific languages embedded in Python. 
The decision procedure uses the OpenFST library~\cite{openfst} and the Python bindings provided by HFST~\cite{hfst} to construct and compose finite state automata and transducers. We implemented certain automaton operations, such as the product relation ($P_1 \times P_2$), ourselves using low-level HFST APIs that manipulate automata directly.

For each flow equivalence class, \sysname reads the before and after forwarding paths from file input, which is produced by the same network simulation toolchain described in \Cref{sec:strawman_manual}.
We tweaked the simulator to output forwarding paths in the \sysname-defined graph format compactly and to enable efficient FSA construction for \lstinline{PreState} and \lstinline{PostState} expressions (\Cref{sec:compilation}). Each equivalent class is processed in parallel.



\para{Practical Extensions}
Each equivalence class specifies the set of IP addresses for the traffic. On occasion, we must specialize analysis to specific IP addresses. To do so, we allow change specifications of the form \emph{prefix-predicate} $\rightarrow$ \emph{change-spec}. 
Semantically, such a change spec is applied exclusively to traffic
classes that satisfy the prefix-predicate.
The predicate language supports filtering based on source and destination IPs and set operations.
%
For example, decommissioning an IP prefix is a common change for which we want to ensure that the network does not carry traffic for these prefixes along any path. We can encode this requirement (for $10.0.0.0/24$) using the following specification.
\begin{lstlisting}
spec dealloc := .* : remove(.*) 
pspec deallocP := 
  (dstPrefix==10.0.0.0/24) -> dealloc
\end{lstlisting}
Such address-based filtering sits outside of the core language and acts as a filter on the forwarding path data.

\section{Case study}
\label{sec:case_study}


We ran \sysname on historical changes in the global backbone. 
Our workflow shared the first four steps with the current workflow in \Cref{sec:strawman_manual}: simulate pre- and post-change networks, compute forwarding paths, aggregate flows into equivalence classes. The final step is different: the forwarding data is given to \sysname as input, along with a spec, and we analyze all flow equivalence classes rather than just the diff. 
We first describe how this process played out for the change in \Cref{sec:example} and then draw lessons from our experience. 



\subsection{Revisiting the example change}
\label{sec:case_study_example}

For each proposed change (i.e., "iteration"), we used \sysname to check the change against
a relational specification.

\para{First iteration} We invoked \sysname with the change implementation v1 (\Cref{fig:shift_exit_v1}) and the \lstinline{change} spec in \Cref{sec:overview}. For this implementation, the path diff of the manual inspection tool had 17 flow equivalence classes. Engineers investigated each class and discovered that none of them corresponded to the desired path change, and all of them stemmed from either issues with the simulation tool or benign side effects of the change. The allow-list change on $A2$ routers caused unexpected but acceptable traffic changes. 

\sysname produced 17 counterexamples for \lstinline{nochange} and 15 for \lstinline{e2e}. The 15 violations for \lstinline{e2e} clearly signaled that the change failed to move T1 traffic, as the pre-change and post-change paths were still the same for such flows. The counterexamples for \lstinline{nochange} are the same as those reported by the path diff tool. To automatically exclude such benign violations in future iterations (and avoid triaging the same warnings again), we extended the spec with a new component called \lstinline{sideEffects}, to explicitly permit such changes.

\para{Second iteration} In the second iteration, we provided the change implementation v2 (\Cref{fig:shift_exit_v2}) and the refined spec. For this implementation, the current path diff tool produced a path diff with 46 classes. 
Engineers waded through them to discover the collateral damage and, because of information overload, missed that the change to T1 traffic was incorrect. 

\sysname produced 15 counterexamples for \lstinline{e2e}, 24 for \lstinline{nochange}and 0 for \lstinline{sideEffects}. The violations signaled that changes to T1 traffic was wrong and there was collateral damage too. The refined \lstinline{sideEffects} spec provided value by suppressing benign differences. 

\para{Final iteration} Because \sysname discovered two errors at the same time, we skipped the third iteration (which was needed during the original manual analysis), and jumped straight to the final iteration. In this iteration, we supply the correct change implementation to \sysname and the refined spec. \sysname validated the change automatically and completely. In contrast, in their original debugging effort, the engineers had to manually inspect the path diff to certify the change. 


\subsection{Lessons learned}

Based on our experience with \sysname, we draw these lessons:

\begin{enumerate}

\item \sysname's categorization of violations based on which sub-spec is violated speeds up error diagnosis and reduces the number of iterations. Errors are quickly diagnosed because the violated sub-spec provides strong hints about their nature; the types of errors that violate \lstinline{nochange} are different from those that violate \lstinline{e2e}. The number of iterations is reduced because multiple errors in an implementation are easier to spot, especially when spread across different sub-specs. With manual inspection, when analyzing a big bag of path diffs, it is hard to spot multiple errors.


\item \sysname specs may need refinement because the original change intent (in natural language) is under-specified or the network is not configured as expected.
Under-specification and unexpected behaviors are common for large networks. However, while the effort put into a manual audit is hard to reuse, effort put into refining a \sysname spec pays off. The refined spec saves work during future iterations of the same change or other similar changes. Multiple changes of the same type are a common occurrence for production networks. 

\item When a change (sub) spec does not match an implementation, there is less data to analyze.  Change implementations are often partially correct, and \sysname produces only violations. The current path diff contains both compliant and non-compliant changes.  The engineers must analyze both to find violations.  


\item When the change spec matches the implementation, the engineers need to do nothing. They can have greater confidence (and peace of mind) in the change compared to manually inspecting the path diff. 

\end{enumerate}

\section{Evaluation}\label{sec:eval}

To evaluate the expressiveness and performance of \sysname, we apply it to a set of real network changes in the global backbone of a large cloud provider. This dataset has all changes that were reviewed by the network's technical committee from Jun 2023 to Jan 2024. The committee reviews all high-risk, complex changes. There are 10s of changes in the dataset; we do not reveal the exact count for confidentiality.

\begin{figure*}[t] 
  \centering
  \begin{minipage}{.26\textwidth}
    \centering
    \includegraphics[width=\linewidth]{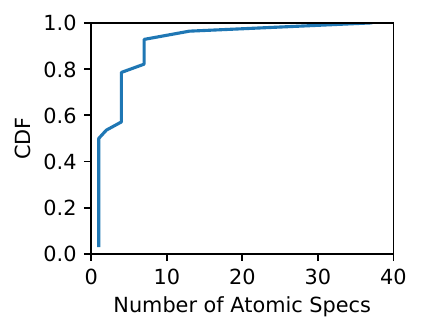}
    \vspace{-20pt}
    \caption{Distribution of spec size in our dataset.}
    \label{fig:eval_loc}
  \end{minipage}%
  \hfill
  \begin{minipage}{.26\textwidth}
    \centering
    \includegraphics[width=\linewidth]{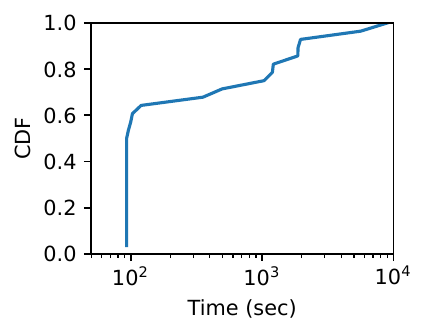}
    \vspace{-20pt}
    \caption{Time (log scale) to validate changes with \sysname.}
    \label{fig:perf_cdf}
  \end{minipage}%
  \hfill
  \begin{minipage}{.4\textwidth}
    \centering
    \includegraphics[width=\linewidth]{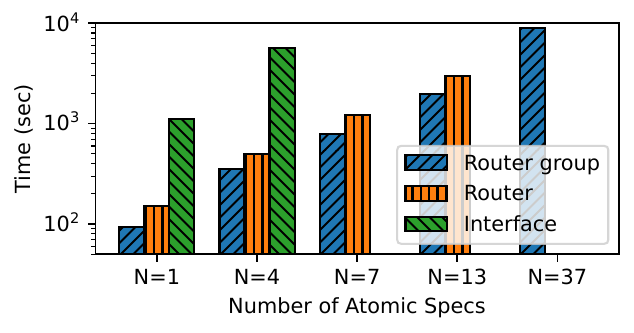}
    \vspace{-20pt}
    \caption{\sysname's validation time (log scale) for different spec sizes and location granularity.}
    \label{fig:perf_microbenchmark} 
  \end{minipage}
\end{figure*}

\subsection{Expressiveness}

We used \sysname to specify engineers' intent for each change in the dataset.
We determined the intent by examining change tickets, which contain a description of the intent in natural language as well as a change implementation plan.  
The tickets describe change intents pertaining to the network data plane as well as 
those of other types, such configuration settings and backup routes. 
We focused on data plane change intents. All changes in our dataset have a data plane change intent; three in four have only data plane change intents.

We found that \sysname can specify the intended data plane change for 97\% of the changes in our dataset. That \sysname can support this high a fraction of high-risk changes in a large, complex network is a highly encouraging result.

For the remaining 3\% of the changes, \sysname could only partially specify the intended data plane change. \sysname's key current limitation is a lack of support for path counting:  In addition to path shape, users sometimes want to limit the number of paths that a flow can take.  For example, because of router hardward limits, one might not want the number of ECMP (equal cost multipath routing) paths for a flow to exceed 128. We will explore supporting such intents in the future by generalizing the \lstinline{any} modifier to include a path count. 

To assess the compactness of specifications, we quantify their size as the number of atomic \sysname specs (of the form \lstinline{r: m}). This analysis excludes any spec refinement that may be needed to accommodate benign side effects (\Cref{sec:case_study_example}); we do not have the data to make that determination. 
\Cref{fig:eval_loc} shows a cumulative distribution function (CDF) of the number of atomic specs needed across all changes. 
The vast majority of the changes (93\%) can be expressed with fewer than 10 atomic specs. The outliers correspond to infrequent, complex changes to the backbone's routing architecture in which significant traffic carried by the network is shifted.

Half the changes require only one atomic spec, corresponding to no expected impact on the forwarding behavior. It may seem odd at first that so many high-risk changes fall in this bucket. But fully preserving forwarding behavior while something is changed under the hood (e.g., modifying the routing policy to replace concrete routes with aggregate routes or standardizing on community tags) is common. It is also high-risk. Indeed, there are changes in our data where no behavior change was expected but the path diff revealed forwarding changes that could have led to an outage.





\subsection{Performance}


We benchmark \sysname's performance by measuring the time to validate changes in our dataset, including the time to deserialize the forwarding path data, FSA/FST construction and equivalence checking. This experiment was done on a computer with 96 CPU cores and 768 GB DRAM. 
Because we did not have access to the precise data plane states of historical changes, we ran all specs on the same data plane state produced by a recent snapshot.

\Cref{fig:perf_cdf} shows a CDF of the validation time. Half of the changes take 93 seconds, which is the time to check the "no change" spec. Four in five changes need less than 20 minutes, and the most complex change needs 150 minutes. For context, we observe that it takes 140 minutes to simulate both network snapshots and compute forwarding paths. We conclude from these results that the performance of \sysname is acceptable for the backbone network, especially considering how long manual inspection takes today.

Diving deeper into \sysname's performance, we find that the two most important factors are the size of the spec (number of atomic specs) and the location granularity. \Cref{fig:perf_microbenchmark} shows this impact by running specs in our data at different granularities. (\Cref{fig:perf_cdf} used the granularity indicated by the change intent, so it has a mix.) 
We exclude granularity-size combinations that need over 3 hours. 

We see that validation time grows with the spec size, and finer granularity analysis takes more time (as expected). The impact of going from the router group level to the router level is small, but the impact of going to the interface level is substantial (10x), due to the substantially higher number of paths at the interface level. Fortunately, under 4\% of the changes in our data require interface-level granularity. 7\% require device-level.

\section{Related Work}

Our work builds on the foundation laid by single-snapshot verification tools~\cite{anteater,hsa,batfish,veriflow,atomic,rcdc,hoyan,minesweeper,arc,era,tiramisu,plankton,netkat}. The application of these tools to real-world networks has improved reliability and provided insights into problem they do and do not solve. We act upon one such insight: that 
many large, real-world networks are difficult to specify accurately in their
entirety.  Without such single-snapshot specifications, engineers need 
different kinds of tools to help them validate network changes automatically. 


Differential network analysis (DNA)~\cite{dna} shares our perspective on network verification---that it is
crucial to track similarities
and differences between pre- and post-change networks. 
It simulates the pair of pre- and post-change control planes efficiently to generate differences in their data plane states. (\sysname makes no
contributions to control plane simulation.) In addition to showing path diffs, DNA can generate differences in single-snapshot invariants, e.g., "A can reach B in the pre-change network but not the post-change network."  Engineers must manually inspect the path and invariant diffs to determine whether or not they indicate errors.  In contrast,
\sysname specifications characterize what constitutes an error,
and our decision procedures check these specifications automatically.
Importantly, \sysname's specifications can be perfectly precise, more precise than "A can reach B"---any specific path
or regular set of paths may be specified. This precision takes manual audits completely out of the loop when changes are conformant.


Batfish supports differential analysis as well~\cite{batfish-differential}. It independently analyzes two snapshots and formats the outputs such that the differences are easier to analyze.  Like DNA, it requires humans to certify correctness and does not have a relational spec.  Once again, \sysname improves on this situation using a relational specification language and deciding the validity of specifications without human auditing.


\sysname was also inspired in part by past work on NetKAT~\cite{netkat}, which has shown that using regular languages (Kleene algebra) is an effective way to specify network behavior.  \sysname builds on ideas from NetKAT by using regular relations in addition to regular languages to express differences and similarities between pairs of networks.

Researchers have explored relational verification for ordinary programs many times in the past~\cite{barthe2011,benton2004,chen2019}. The archetypal goal here is to verify that two programs are equivalent. At least superficially, the techniques for relational program verification differ from those in \sysname. A common method is to consider a "product" program that combines two input programs and verify the safety properties of this product. An interesting avenue for future work is to consider whether specific relation program verification techniques can help us verify networks more efficiently or vice versa.

\section{Summary}
\label{sec:conclusions}

We develop the concept of relational network verification and realize it in the \sysname tool for validating network changes. Our key observation is that relational specifications can compactly and precisely capture change intents; they need only express what is expected to change, which is often a miniscule fraction of the overall network, and simply say "no change" for the rest. 
For a global backbone with over $10^3$ routers, 93\% of the high-risk changes need fewer than 10 terms and 80\% of them can be validated in under 20 minutes. 
We look forward to exploring applications of relational methods to other network verification and synthesis problems in the future.

\balance
\bibliographystyle{ACM-Reference-Format}
\bibliography{reference}

\newpage

\appendix
\section{Semantics of \sysname RIR} \label{sec:semantics}


The semantics of $Path\ Set$ is given by equations of the form
$\mathscr{P} \llbracket P \rrbracket (M, N) \triangleq S$ where
$M$ and $N$ are sets of paths representing the old and new networks
respectively and $S$ is the resultant set of paths denoting $P$.

\begin{align*}
\mathscr{P}\llbracket a \rrbracket (M, N) &\triangleq \{a\} \\
\mathscr{P}\llbracket 0 \rrbracket (M, N) &\triangleq \emptyset \\
\mathscr{P}\llbracket 1 \rrbracket (M, N) &\triangleq \{\epsilon\} \\
\mathscr{P}\llbracket \mathtt{PreState} \rrbracket (M, N) &\triangleq M \\
\mathscr{P}\llbracket \mathtt{PostState} \rrbracket (M, N) &\triangleq N \\
\mathscr{P}\llbracket P_1\mid P_2 \rrbracket (M, N) &\triangleq\mathscr{P}\llbracket P_1\rrbracket (M, N) \cup \mathscr{P}\llbracket P_2 \rrbracket (M, N) \\
\mathscr{P}\llbracket P_1P_2 \rrbracket (M, N) &\triangleq \{p_1p_2 \mid p_1 \in \mathscr{P}\llbracket P_1\rrbracket (M, N),\\&p_2 \in \mathscr{P}\llbracket P_2 \rrbracket (M, N) \} \\
\mathscr{P}\llbracket P^* \rrbracket (M, N) &\triangleq \{p_1...p_n \mid p_1,...,p_n \in \mathscr{P}\llbracket P\rrbracket (M, N)\} \\
\mathscr{P}\llbracket P_1 \cap P_2 \rrbracket (M, N) &\triangleq\mathscr{P}\llbracket P_1\rrbracket (M, N) \cap \mathscr{P}\llbracket P_2 \rrbracket (M, N) \\
\mathscr{P}\llbracket \overline{P} \rrbracket (M, N) &\triangleq \Sigma^* \backslash \mathscr{P}\llbracket P_1\rrbracket (M, N)\\
\mathscr{P}\llbracket P \triangleright R \rrbracket (M, N) &\triangleq \{q \mid \exists p.~\langle p,q \rangle \in \mathscr{R}\llbracket R \rrbracket (M, N) \\&\wedge p \in \mathscr{P}\llbracket P \rrbracket (M, N) \} \\
\end{align*}

The semantics of $Rel$ is given by equations of the form
$\mathscr{R} \llbracket R \rrbracket (M, N) \triangleq T$ where
$M$ and $N$ are sets of paths representing the old and new networks
respectively and $T$ is the resultant set of pairs of paths denoting
relation $R$.

\begin{align*}
\mathscr{R}\llbracket P_1 \times P_2 \rrbracket (M, N) &\triangleq \{\langle p_1, p_2 \rangle \mid p_1 \in \mathscr{P}\llbracket P_1\rrbracket (M, N),\\&p_2 \in \mathscr{P}\llbracket P_2\rrbracket (M, N)\} \\
\mathscr{R}\llbracket 0 \rrbracket (M, N) &\triangleq \emptyset \\
\mathscr{R}\llbracket 1 \rrbracket (M, N) &\triangleq \{(\epsilon,\epsilon)\} \\
\mathscr{R}\llbracket R_1\mid R_2 \rrbracket (M, N) &\triangleq\mathscr{R}\llbracket R_1\rrbracket (M, N) \cup \mathscr{R}\llbracket R_2 \rrbracket (M, N) \\
\mathscr{R}\llbracket \mathtt{I}(P) \rrbracket (M, N) &\triangleq \{\langle p, p \rangle \mid p \in \mathscr{P}\llbracket P\rrbracket (M, N)\}\\
\mathscr{R}\llbracket R_1R_2 \rrbracket (M, N) &\triangleq \{\langle p_1p_2,q_1q_2 \rangle \mid \langle p_1, q_1 \rangle \in \mathscr{R}\llbracket R_1\rrbracket (M, N),\\& \langle p_2, q_2 \rangle \in \mathscr{R}\llbracket R_2\rrbracket (M, N)\} \\
\mathscr{R}\llbracket R^* \rrbracket (M, N) &\triangleq \{ \langle p_1...p_n, q_1...q_n \rangle \mid \\&  \langle p_1, q_1 \rangle, ..., \langle p_n, q_n \rangle \in \mathscr{R}\llbracket R\rrbracket (M, N)\} \\
\end{align*}

The semantics of $Spec$ is given by a satisfaction relation $M, N \models S \Longleftrightarrow Bool$ where 
$M$ and $N$ are sets of paths representing the old and new networks
respectively.   These sets of paths satisfy the spec $S$ exactly when the right-hand-side is true.

\begin{align*}
M, N \models P_1 = P_2 &\Longleftrightarrow \mathscr{R}\llbracket P_1 \rrbracket (M, N) = \mathscr{R}\llbracket P_2 \rrbracket (M, N) \\
M, N \models P_1 \subseteq P_2 &\Longleftrightarrow \mathscr{R}\llbracket P_1 \rrbracket (M, N) \subseteq \mathscr{R}\llbracket P_2 \rrbracket (M, N) \\
M, N \models S_1 \wedge S_2 &\Longleftrightarrow M,N \models S_1 ~\text{and}~ M,N \models S_2 \\
M, N \models S_1 \vee S_2 &\Longleftrightarrow M,N \models S_1 ~\text{or}~ M,N \models S_2 \\
M, N \models \neg S &\Longleftrightarrow M,N \not\models S 
\end{align*}

\end{document}